\documentclass[12pt]{article}

\usepackage{amsmath}
\usepackage{amsfonts}
\usepackage{amssymb, amsthm}

\begin{document}


\title{A Survey of Gravitational Waves}

\author{Frans Pretorius\\
        Department of Physics\\ 
        Princeton University\\
        Princeton, NJ 08544\\
        fpretori@princeton.edu}



\begin{abstract}
We review the state of the field of gravitational wave astrophysics,
framing the challenges, current observations, and future prospects within the
context of the predictions of Einstein's theory of general relativity.

\end{abstract}

\date{January 2022}
\maketitle


\section{Introduction}
This article\footnote{To appear in the International Congress of Mathematicians 2022 (ICM 2022) Proceedings.} 
is meant to serve as an overview of the current state of the field
of gravitational wave astrophysics. It is not meant to be comprehensive, nor a reference
for experts, but rather an introduction to this nascent field of observational
science, targeted toward mathematicians and scientists. The three primary
goals are (a) to give a sufficient introduction to the physics of general
relativity to appreciate the challenges of gravitational wave detection,
as well as the remarkable nature of sources of gravitational
waves in the dynamical, strong field regime of the theory,
(b) to review what has been learnt about the Universe
from the gravitational wave signals detected to date by
the LIGO (Laser Interferometer Gravitational-Wave Observatory)/Virgo detectors,
and (c) to briefly speculate about future discoveries that will unfold over the coming decades
as a variety of observational campaigns are undertaken.
To set the stage then, in Sec.~\ref{sec_einstein} we review the underlying theoretical
framework, Einstein's theory of general relativity, focusing
on the nature of gravitational waves and how they are produced.
In Sec.~\ref{sec_obs} we briefly survey the current detectors and observational campaigns,
either in operation today or planned for the
coming decade or two : ground based detectors (as LIGO/Virgo), the space based mission
LISA (Laser Interferometer Space Antenna), pulsar timing arrays, and the search for B-mode
polarization of the cosmic microwave background (CMB).

LIGO measured the first
gravitational wave signal, GW150914, in 2015, which is interpreted as originating from 
the merger of two black holes~\cite{LIGOScientific:2016aoc}. Since
then, LIGO/Virgo has observed almost 100 additional signals, most also from
binary black hole mergers, though a small handful likely coming from
black hole/neutron star or binary neutron star mergers. However the loudest event
to date, GW170817, was a binary neutron star merger, as confirmed
by a spectacular suite of electromagnetic observations of its aftermath.
In Sec.~\ref{sec_todate} we review these observations, and what they have so far
taught us about the Universe. Highlights are the first quantitative
evidence that black holes as described by Einstein's theory do in fact exist, that
the speed of gravitational waves is equal to that of the speed of
light to within $\sim 1$ part in $10^{15}$, and that
neutron star mergers are responsible for at least a class of the mysterious
so-called short gamma ray bursts (observed at a rate of about one every 3 days
by special purpose satellites designed for this).

We conclude in Sec.~\ref{sec_future} with speculations on the coming two decades
of gravitational wave astronomy. 

\section{Einstein Gravity}\label{sec_einstein}
The working hypothesis upon which the science of gravitational wave astrophysics
is built is that ``gravity'' is described by Einstein's classical theory
of general relativity. This begins by positing that space and time taken together, or spacetime for short,
has the structure of a 4-dimensional Lorentzian geometry. A convenient way to describe
this geometry is via the metric tensor $g_{ab}$, defined in a coordinate
basis through the line element
\begin{equation}
ds^2 = g_{ab} dx^a dx^b,
\end{equation}
which gives the local, infinitesimal proper distance-squared $ds^2$ as a quadratic
form of an arbitrary infinitesimal coordinate displacement $dx^a$ (we use
the Einstein summation convention where repeat indices in a tensor expression
implies summation). The phrase {\it proper distance} means
the coordinate invariant, physically
measurable length or time interval, in contrast to a coordinate distance
in some (arbitrary) coordinate system. 
The Lorentzian (indefinite) character of the metric is crucial,
as it allows one to define causality through geometry: two different events
are causally related if and only if there exists at least one curve connecting them where
the proper distance along the curve is everywhere timelike, 
$ds^2<0$, and/or null, $ds^2=0$ (the sign convention for
timelike $ds^2<0$ versus spacelike $ds^2>0$ is arbitrary).

The second key postulate of general relativity is that the geometry
of spacetime relevant to the Universe is not a fixed structure given {\it a priori}, but
instead is a dynamical entity governed by the Einstein field equations:
\begin{equation}\label{efe}
G_{ab} \equiv R_{ab} - \frac{1}{2} R g_{ab} = \frac{8\pi G}{c^4} T_{ab},
\end{equation}
where the Einstein tensor $G_{ab}$ is defined as above in terms
of the Ricci tensor $R_{ab}$ and Ricci scalar $R$, $T_{ab}$ is the
stress-energy-momentum tensor of the matter content of the Universe,
$G$ is Newton's constant and $c$ is the speed of light. General
relativity ignores torsion, which is thought would only be needed
to describe matter with intrinsic spin, and is expected to be
irrelevant for macroscopic distributions of matter in the classical limit.
Thus all tensors appearing in (\ref{efe}) are symmetric.
In terms of practically solving this equation,
one views the Einstein tensor as a second order, quasi-linear partial
differential operator acting on the metric tensor $g_{ab}$. In 4 spacetime
dimensions, this gives a set
of 10 coupled equations for the independent components
of $g_{ab}$, and must be solved simultaneously with the additional equations governing the matter fields
in $T_{ab}$. It is obvious
from (\ref{efe}) then that matter ($T_{ab}$) will influence the dynamics
and curvature of spacetime. Less obvious is even in the absence
of matter ($T_{ab}=0$) non-trivial, dynamical solutions exist: most interesting
among these are those describing black holes and gravitational waves.

It is often stated that a third key postulate of general relativity is
the {\it geodesic hypothesis} : a test body not subject to any force 
follows a geodesic of the spacetime (a test body is one
with insufficient energy to cause any noticeable perturbation on the surrounding
geometry). However, perhaps more fundamentally, geodesic motion
in the test body limit can be viewed as coming from 
energy/momentum conservation, which is already built into the Einstein
equations and does not need to be imposed as a separate hypotheses.
This follows from the contracted Bianchi identities, showing
that the Einstein tensor necessarily has vanishing divergence $\nabla_a G^{ab}=0$.
Thus, any matter that can self-consistently be coupled to spacetime through the 
Einstein equations (\ref{efe}) must have a divergenceless stress tensor $\nabla_a T^{ab}=0$,
the latter equation being the covariant statement of the conservation of energy/momentum of the matter.
Likewise, pure spacetime energy, whether in the form of gravitational
waves, or confined to black holes, will exhibit similar dynamics
in an equivalent test body limit. For example, in vacuum an infinitesimal mass black hole
will orbit a large (finite mass) black hole following a geodesic of the latter's
spacetime by virtue of the vacuum Einstein equations alone, and not any additional
hypothesis one needs to supply.

If the nature of spacetime is as described by general relativity,
the most immediate consequences of this are well described by Newtonian's theory of gravity 
in the weak field limit (for example our environment here on Earth and in the solar system).
This is why Einstein's theory is also called a theory of gravity despite there
being no gravitational force in general relativity.

\subsection{Gravitational Waves}

It is not possible to precisely define what a gravitational wave is in
all scenarios. For our purposes, it suffices to think of gravitational
waves as small, local disturbances in spacetime that
propagate at the speed of light. In an asymptotically flat space time
(the metric at large distances from any source of curvature approaches that of
special relativity---Minkowski spacetime)
the properties of gravitational waves can be defined more precisely.
Our Universe is {\it not} asymptotically flat, though with appropriate
accommodation for the overall cosmic expansion with time, to good
approximation we can consider ourselves to be in an asymptotically
flat region relative to any source we expect to observe.

Regarding sources of gravitational waves, there are two broad classes. First is what
one traditionally thinks of as a source : at some place a localized event
occurs that produces gravitational waves over a period of time, and
these waves then stream outward away from the source.
Second are ``primordial'' gravitational waves, namely an overall background of 
gravitational waves filling all of space, having been produced in an earlier epoch of the 
evolution of the Universe.
In some cases the distinction between these classes is blurred; for
example, a sufficiently high density of localized sources emitting over
a long period of time will eventually also fill the observable Universe
with a background of gravitational waves.
In these settings then, we next review some of the basic properties
of gravitational waves, and how they are produced.

\subsubsection{Basic Properties of Gravitational Waves in the Weak Field Limit}\label{sec_gws}
Consider a metric perturbation $h_{ab}$ about a background Minkowski spacetime
$\eta_{ab}$, i.e. $g_{ab}=\eta_{ab}+h_{ab}$ with  
$\eta_{ab}dx^a dx^b=-c^2 dt^2+dx^2+dy^2+dz^2$ in Cartesian coordinates.
Then the linearized Einstein
equations show that general relativity allows two linearly independent
gravitational wave solutions for $h_{ab}$, or so-called {\it polarizations}\footnote{In principle
a general metric theory of gravity can allow up to 6 linearly
independent polarizations; see e.g.~\cite{Will:2014kxa}.}, propagating in any given direction.
Even restricting the background metric to be in Cartesian form, there
is still much coordinate (or ``gauge'') freedom to choose the representation
of the solution. A gauge commonly used is the so-called {\it transverse traceless} gauge,
and in these coordinates a wave propagating in the $+z$ direction (for example) 
takes the form (e.g.~\cite{Flanagan:2005yc,Buonanno:2007yg})
\begin{equation}\label{tt_wave}
h_{ab} dx^a dx^b=h_+(t-z/c)[dx^2-dy^2] + h_\times(t-z/c)[2dxdy].
\end{equation}
$h_+$ and $h_\times$ are arbitrary (but small amplitude) functions of their
arguments, and describe the so-called {\it plus} and {\it cross} polarized
waves respectively. From (\ref{tt_wave}) one can see that gravitational
waves in general relativity are transverse, namely they only perturb
the background metric along a plane (the $x,y$ plane in this example) orthogonal
to the direction of propagation ($z$ here). 
Equation (\ref{tt_wave}) also shows that as a plus polarized
wave passes a given point, when $h_+>0$ it will stretch proper
distances in $x$ by $\sqrt{1+h_+}$ while simultaneously
squeezing distances in $y$ by $\sqrt{1-h_+}$, and the opposite when $h_+<0$.
The effect of the cross polarized wave on the transverse geometry
is qualitatively the same, except the directions of stretching/squeezing
are rotated by $45^\circ$ about the $z$ axis relative to that of the plus polarized wave.

The energy flux density carried by these waves is
\begin{equation}\label{gw_e}
\frac{dE}{dA dt} = \frac{c^3}{16\pi G}\left<\left(\frac{dh_+}{dt}\right)^2  + 
                                              \left(\frac{dh_\times}{dt}\right)^2 \right>,
\end{equation}
where $dA$ is the transverse area element, and the angle brackets
denote a time average over a characteristic period of the wave (the reason
for the averaging is that gravitational wave energy cannot be localized---see e.g.~\cite{Misner:1973prb}).
A truly infinite plane wave such as (\ref{tt_wave})
will have infinite total energy, which is not consistent with an asymptotically flat space time
when backreaction is taken into account.
However, sufficiently far from a local source (as discussed below) the outgoing spherical wavefronts
are locally well approximated by these plane wave solutions. Similarly, an on-average
homogeneous, primordial stochastic background that fills all of spacetime can 
not be asymptotically flat when backreaction is considered\footnote{Instead then one obtains
the Friedmann-Robertson-Lema\^itre-Walker (FRLW) asymptotics that observations
indicate describe our Universe on very large scales.}, but still the above (generalizing to superpositions
of plane waves travelling in all directions) can give a good description of the geometry
in any local patch of the spacetime.

Notice the way $G$ and $c$ appear in (\ref{gw_e}), and hence the dimension-full constant relating
energy flux on the left hand side to the time derivative of metric strain on the right hand side: 
in SI units $c^3/G\sim 10^{36}\rm{J\cdot s/m^2}$. 
This implies, at least from the perspective of our everyday intuition of energy
and length scales, that it requires an enormous amount of energy to perturb
spacetime by a comparatively miniscule amount. This is the reason why it is completely
impractical to study gravitational waves by building transmitters/receivers on earth
in analogy with electromagnetic waves. Instead we must look to cataclysmic 
gravitational wave ``explosions'' in the cosmos, such as those produced by black hole mergers, 
and even
then, despite the astonishing sensitivity of the LIGO/Virgo detectors, we are now just
barely able to observe them.

Regarding localized sources of gravitational waves, good insight can again
be obtained from linearized theory, resulting in the so-called quadrupole
formula. Here, one assumes a weak field, slowly varying distribution
of energy density $\rho(t,x,y,z)$. This will emit gravitational waves propagating 
outward that at a large distance $r$ from the source takes the following form 
in terms of the spatial components of the metric perturbation $h_{ij}$:
\begin{equation}\label{quad}
h_{ij}(t,r) = \frac{1}{r}\frac{2 G}{ c^4 } \frac{d^2\mathcal{I}_{kl}(t-r/c)}{dt^2}
         \left[P_i{}^k P_j{}^l - \frac{1}{2} P^{kl} P_{ij}\right],
\end{equation}
with all indices here only running over the spatial coordinates $x_i\in (x,y,z)$ 
(in transverse traceless gauge
there are no time-time or space-time propagating components of $h_{ab}$),
$\mathcal{I}_{ij}$ is the reduced quadrupole moment tensor
of the source, and the projection
tensor $P_{ij}\equiv \delta_{ij} - n_i n_j$, where $n^i$ is a unit spatial
vector pointing from the origin $r=0$ to the observer location $r=\sqrt{x^2+y^2+z^2}$ .
$\mathcal{I}_{ij}$ is defined in terms of the quadrupole moment tensor $I_{ij}$  as
\begin{equation}
\mathcal{I}_{ij} \equiv I_{ij} - \frac{1}{3} I^k{}_k \delta_{ij}, \ \ \ I_{ij}(t)\equiv \int x_i x_j \rho(t,x,y,z) dV,
\end{equation}
where the integration is over all of space at some instant of time $t$, but note that 
in deriving (\ref{quad}) the source
is assumed to be localized in space around $r=0$, and the observer 
location $r\gg 0$ is assumed to be in vacuum.

Several properties of gravitational wave emission
are evident from (\ref{quad}). First, unsurprisingly, the
outgoing wave propagates at the speed of light, and its amplitude
decays with distance like $1/r$ from the source. Second, similar to that implied by the
energy expression in (\ref{gw_e}), the factor of $G/c^4\sim 10^{-44} \rm{s^2/kg/m}$ 
illustrates what extreme dynamics,
in the form of rapid accelerations of large energy densities,
need to be present in the source
to produce non-negligible metric perturbations. Third, it
is only the acceleration of {\it asymmetric} concentrations of
energy that produce gravitational waves in general relativity; for example,
a spherically symmetric pulsating star cannot produce any gravitational waves.

\subsubsection{Weak Field Emission from a Compact Object Binary}

Though it is not obvious from the discussion above, it turns
out that the quadrupole formula (\ref{quad}) gives a good approximation
to the gravitational wave emission even for certain strong field sources,
and even if the energy density is purely gravitational, such as with black hole
binaries. Another property of binary systems
in general relativity that we will simply mention without giving further details, is that
backreaction from the loss of energy to gravitational wave emission 
not only causes the semi-major axis of the binary to decrease (as anticipated
by Newtonian energy balance), but it also reduces the eccentricity of the binary
with time. LIGO/Virgo is only sensitive to the very last stages
of binary inspiral, and the majority of observable systems are thus expected to 
have close to zero eccentricity. In all then,
to get a good understanding of the structure of gravitational waves
emitted by such a so-called quasi-circular inspiral, we can 
evaluate the quadrupole formula (\ref{quad}) for two
point masses $m_1$ and $m_2$ orbiting each other on a circle
separated by a distance $D$, with orbital frequency $\omega$, which for large
separations is well approximated by the Keplerian result $\omega=\sqrt{G M/D^3}$, with $M=m_1+m_2$.
For a binary orbiting in the $z=0$ plane about $r=0$, using spherical
polar coordinates to label the observer location $(x,y,z)=(r\cos\phi\sin\theta,r\sin\phi\sin\theta,r\cos\theta)$,
and expressing the answer in terms of the two polarization amplitudes
in the plane orthogonal to the propagation vector $n^i$, gives
\begin{eqnarray}
h_+(t,r,\theta,\phi)&=&\frac{1}{r}\frac{4 G}{c^4}\mu D^2 \omega^2\cos(2\omega(t-r/c)-2\phi)
\left[\frac{1+\cos^2\theta}{2}\right]\label{hp},\\
h_\times(t,r,\theta,\phi)&=&\frac{1}{r}\frac{4 G}{c^4}\mu D^2 \omega^2\sin(2\omega(t-r/c)-2\phi)\cos\theta\label{hx},
\end{eqnarray}
where $\mu=m_1 m_2/(m_1+m_2)$ is the reduced mass of the binary, and an arbitrary initial phase was set to zero.
The corresponding orbit averaged energy fluxes, from (\ref{gw_e}), are
\begin{eqnarray}
\frac{dE_+}{dAdt} &=& \frac{2}{\pi r^2}\frac{G}{c^5}\mu^2 D^4 \omega^6 
\left[\frac{1+\cos^2\theta}{2}\right]^2\label{hpe},\\
\frac{dE_\times}{dAdt} &=& \frac{2}{\pi r^2}\frac{G}{c^5}\mu^2 D^4 \omega^6 \cos^2\theta\label{hxe}.
\end{eqnarray}
Integrating these over the sphere gives the net radiated power in the two modes
\begin{eqnarray}
\frac{dE_+}{dt} &=& \frac{56}{15}\frac{G}{c^5}\mu^2 D^4 \omega^6\label{hpte},\\ 
\frac{dE_\times}{dt} &=& \frac{8}{3}\frac{G}{c^5}\mu^2 D^4 \omega^6\label{hxte}.
\end{eqnarray}
Several interesting properties are apparent from expressions (\ref{hp}-\ref{hxte}) :
the observed gravitational wave frequency is twice the orbital frequency, 
the orbit averaged 
amplitudes (hence energy fluxes) are not isotropic in latitude,
nor is emission equally balanced between the plus
and cross polarizations. Also, as expected, the emission vanishes
in the test body limit $\mu\rightarrow 0$. The total energy flux $dE/dt=32 G \mu^2 D^4 \omega^6/5 c^5$,
or using the Kepler relation for $\omega(D)$, is
\begin{equation}\label{gw_lum}
\frac{dE}{dt}=\frac{32 G^4 M^3 \mu^2}{5 c^5 D^5} = \frac{32 G^{7/3} M^{4/3} \omega^{10/3} \mu^2}{5 c^5}. 
\end{equation}
This illustrates how sensitive the luminosity is to orbital separation $D$ or frequency $\omega$.

Note again that equations (\ref{hp}-\ref{gw_lum}) do {\it not} include back reaction; we have
simply evaluated the quadrupole formula for two point masses moving in a circular orbit. 
To obtain the so-called Newtonian quasi-circular approximation to estimate the radiation
reaction on the orbit, one elevates the frequency (or equivalently separation) to a function of time $\omega(t)$,
assumes the Newtonian expression for the energy of the orbit, and uses the latter
together with the total luminosity of the binary to derive an equation 
for the evolution of $\omega(t)$ consistent
with total energy conservation. The result is 
\begin{equation}\label{wt}
\omega^{-11/3}\frac{d\omega}{dt}= \frac{96}{5}\nu\left(\frac{GM}{c^3}\right)^{5/3},
\end{equation}
where $\nu=\mu/M$ is the symmetric mass ratio of the binary.
It is essentially a measurement of (\ref{wt}) from the famous Hulse-Taylor binary
pulsar that gave the first (indirect) evidence for the existence of gravitational
waves, and that the weak-field description of the emission process is consistent with
general relativity. 

\subsubsection{Strong Field Gravitational Wave Emission}\label{sec_SF}
In contrast to the other fundamental laws of physics, the strongly interacting,
or strong field regime of {\it classical} general relativity is not associated
with any particular scale within the theory. Or said another way,
general relativity is a geometric theory, but there is no fundamental
constant of dimension length in the field equations that would
describe a radius of curvature to demark a scale where a qualitative change
in the character of solutions might occur. Despite that, general relativity
{\it does} have a strong field regime, essentially because the field equations
are non-linear. In contrast, Newtonian gravity, a scale-free linear theory, does
not have a strong field regime : the Newtonian gravitational force
can certainly be ``strong'', but it is not qualitatively different 
from a ``weak'' Newtonian gravitational force---they only differ in magnitude.

In general relativity there is no universal criteria for when non-linear
effects become significant enough to qualitatively change solutions,
though for spherical-like compact objects in asymptotically flat spacetime 
there is a good heuristic understanding :
if an amount of energy $M c^2$ is confined to a region 
within a radius (roughly) smaller than its so-called Schwarzschild radius $R_s = 2 G M/c^2$,
the geometry of spacetime qualitatively changes character compared
to a less compact distribution of energy. In particular, spacetime
necessarily becomes dynamical, undergoing what is called {\it gravitational collapse},
and some kind of spacetime singularity forms in the interior. 
A version of Penrose's cosmic censorship conjecture argues that 
generically one expects an event horizon to form about the collapsing
region of spacetime~\cite{1969NCimR...1..252P}; i.e. from an exterior observer's perspective a black hole forms.
If the collapsing region is much more elongated (more cylindrical rather 
than spherical), Thorne's hoop conjecture argues a naked singularity would
form instead~\cite{ThorneHC}, though there are comparatively few studies of such asymmetric
collapse, nor indications that such scenarios arise in astrophysical
settings.

Regarding sources of gravitational waves, again it is not
easy to define when we are in the strong versus weak field
emission regime, though for binary inspiral we can heuristically
characterize the differences. In the weak field the linearized
results described in the previous section are quite accurate.
Somewhat surprisingly, as mentioned, the weak field description
can still be good even if the individual members of the binary by themselves
require strong field gravity to describe their local geometries (case in point
the Hulse-Taylor binary pulsar, as a neutron star's radius
is only a factor of 3 or so larger than its Schwarzschild radius).
A strong field description for a compact binary interaction is necessary either 
when the two objects come close enough that the local geometry of one object
is significantly perturbed by the other (i.e. the point mass approximation
breaks down), 
or the metric perturbation $h_{ab}$ of the observed radiation, when ``scaled back''
by $r$ to the location of the source, becomes of order unity.

Interestingly, the radiative perturbation reaching of order unity
coincides with the gravitational wave luminosity approaching
the Planck luminosity, $L_p=c^5/G$. 
Planck units are a set of units based on the dimensionful constants one can obtain
from the simplest products of powers of the fundamental constants of nature,
in particular $G$, $c$, Planck's constant $h$ and the Boltzmann constant $k$. 
It is theorized that
``quantum gravity'' effects become important when any relevant physical
scale in a process becomes of order unity when measured in Planck units.
The Planck luminosity does not involve Planck's constant, the hallmark
of quantum processes,
yet still, exceeding $L_p$ in a local interaction does seem to anticipate evolution to a regime
where quantum gravity would be necessary.
The reason is based on dimensional
analysis, together with the above heuristic for when one
expects gravity to be so strong that a black hole would
be present, as follows\footnote{To our knowledge arguments like
this were first proposed by Dyson in thought experiments
on whether a single ``graviton'', the hypothetical
quantum particle of geometry, could be detected~\cite{Dyson:2013hbl}.}.
Consider a causal process confined to a volume of characteristic
size $2R$, emitting gravitational waves with total energy $E$. 
For the gravitational waves by themselves to not have enough
energy to form a black hole requires $R$ to be larger than
the effective Schwarzschild radius $2 G E/c^4$ of the gravitational
wave energy, or $E < c^4 R/2G$. If not confined to a black hole,
these gravitational waves will leak out on a light crossing time 
of the system $T=2R/c$, implying a luminosity limit of $L=E/T<L_p/4$.
Or conversely, a process emitting at super-Planck luminosity is necessarily
confined to a black hole, hence censored from exterior observation, 
and whose interior would require some form of quantum gravity for a
complete description.

\subsubsection{Strong Field emission from a Compact Object Binary}
For a quasi-circular binary black hole merger, the weak field description breaks down
primarily because of finite size effects (the two horizons fuse together), and less so because of high gravitational
wave luminosity, which ``only'' reaches up to around $10^{-3} L_p$ for equal
mass mergers ($\mu=M/4$) as computed via full numerical 
solutions~\cite{Pretorius:2005gq}\footnote{A binary neutron star merger has a peak
luminosity a couple of orders of magnitude lower than that of a binary black hole merger.
Finite size effects are more pronounced for neutron stars at late stages
of the inspiral due to their higher tidal deformability, and of course
when they finally collide the point mass approximation used in (\ref{hp}-\ref{gw_lum})
completely breaks down. If the neutron star does not promptly collapse
to a black hole, the gravitational wave emission of the remnant
can still qualitatively be understood using weak field/quadrupole-formula
type analysis, though the complicated dynamics of the matter in the remnant
would not be easy to compute without a numerical solution.}.
To put this number in context, the sun's luminosity in light is $~10^{-26}L_p$;
thus, for the brief moment about the time of merger, a binary black hole radiates
as much power in gravitational waves as $10^{23}$ suns do in light---that is
comparable to the current estimated luminosity of {\it all} stars in the visible Universe combined.
That gravitational wave energy liberated from black hole mergers does not dominate
the energy content of the Universe is in part because they are so rare,
and in part because this incredible luminosity only lasts for a short
time. For example, with GW150914, the merger of two black holes each
roughly $30$ times the mass of the sun $M_\odot$,
the luminosity integrated
over the entire inspiral and merger came to about $3 M_\odot c^2$; the majority
of this was emitted within a few tens of milliseconds~\cite{LIGOScientific:2016aoc}.

The quadrupole formula based
calculation (\ref{gw_lum}) does a decent job of anticipating these
properties, both the rapid increase in luminosity approaching merger,
and the ballpark maximum, if for the latter
we take some liberty in interpreting when the inspiral should terminate.
Rewriting the distance $D$ between the two point masses in (\ref{gw_lum})  
as a fraction $D_s$ 
of the Schwarzschild radius $2GM/c^2$ of the combined mass $M$ of the system, i.e.
$D=D_s\cdot2GM/c^2$, for the equal mass $\mu=M/4$ case gives
\begin{equation}\label{gw_lump}
\frac{dE}{dt}=\frac{L_p}{80 D_s^5}.
\end{equation}
Clearly the maximum inspiral
luminosity depends quite sensitively on $D_s$. For an upper limit,
one would not expect this to
be remotely accurately if $D_s<1$, as then the two horizons of the individual
black holes would already be overlapping. With $D_s\sim 1$, $dE/dt\sim 10^{-2} L_p$.
For a lower limit estimate,  
one can appeal to a result from circular geodesics, where the inner
most stable orbit is at $R=3 R_s$, and then a small loss of angular momentum will
cause the geodesic to plunge into the black hole. Setting $D_s\sim 3$ for the maximum
in (\ref{gw_lump}) thus
amounts to assuming that for comparable mass mergers a similar instability
sets in that accelerates the merger beyond what radiation reaction
does by itself; this gives $dE/dt\sim 10^{-4} L_p$.

Of course, for these back-of-the-envelope estimates to have any relevance
to the maximum merger luminosity requires that the actual collision
of two black holes is not much more violent than the last stages
of inspiral. In fact, before numerical solutions become available, it was
unknown whether black hole collisions would generically even adhere to cosmic censorship,
let alone how bright they ultimately were.
If a merger does satisfy cosmic censorship, the no-bifurcation theorem of Hawking would apply,
telling as two black holes must fuse into a larger one~\cite{Hawking:1971vc}; then also,
by Hawking's area theorem~\cite{Hawking:1971tu}, one can place limits on the maximum amount
of energy that could be liberated in this most non-linear phase of the merger.
If a naked singularity is produced, classical general relativity will
cease to predict the spacetime to the causal future of this event,
and we would have no idea what the remnant of such a black hole collision is. 
Fortunately for our ability to predict waveforms to interpret LIGO events,
but unfortunately for our ability to use black hole mergers to give an observational
glimpse into the mystery of quantum gravity, there is no example yet
from a merger simulation that shows any violation of cosmic censorship,
or anomalously large curvatures forming exterior to the existing
horizons.\footnote{In spacetime dimensions above four there are examples
of (apparent) naked singularity formation from fragmentation of unstable horizons~\cite{Lehner:2010pn},
and hints that certain collisions may also lead to naked singularities~\cite{Okawa:2011fv}.
Though the kind of microscopic extra dimensions that could exist while still 
evading experimental detection will not cause instabilities in astrophysically
sized black holes, and then the effective four dimensional simulations used
to study the latter should be quite accurate.}

Though likely not relevant to the kind of black hole mergers that occur in the
Universe, their is a regime of the two body problem where it {\it is} large gravitational
energy that pushes the interaction to the non-linear regime, and not any finite size
effects : the ultra-relativistic scattering problem.  
Here, one imagines shooting two black holes toward each other at very high
velocities, so that in the center of mass frame of the interaction the kinetic
energy of either black hole is much greater than its rest mass
energy : $(\gamma_i-1) m_i c^2\gg m_i c^2$, with $\gamma_i=1/\sqrt{1-v_i^2/c^2}$.
Though few detailed results are available for the case with generic impact parameter $b$,
it is expected that when $b$ is of order a few times or less than that of the Schwarzschild radius 
$R_s=2G E/c^4=2 G(\gamma_1 m_1 + \gamma_2 m_2)/c^2$ of the system (and note that this scale is much larger
than the Schwarzschild radii of either black hole when $\gamma_i\gg1$), a 
sizable fraction of the kinetic energy can be converted to gravitational
wave energy on a time scale $R_s/c$. Moreover, for $b\lesssim R_s$, an encompassing black
hole forms, trapping most of the kinetic/gravitational wave energy. Again,
exactly how much is not known for generic $b$, though for $b=0$ numerical
simulations show $\sim 14\%$ of $E$ is liberated as gravitational wave energy, with the
remainder trapped~\cite{Sperhake:2008ga}. It has been conjectured that the highest luminosity will be reached
at the critical impact parameter $b_{crit}$ marking the threshold of formation
of a central black hole (for larger impact parameters the two black holes
will fly apart again)~\cite{Pretorius:2007jn}. Then, essentially all of the kinetic energy ($\approx E$)
is expected to be converted to gravitational wave energy, though due to
how strongly this seems to be focused inward when produced, only about
half of this energy may likely escape as gravitational waves~\cite{Sperhake:2012me,Gundlach:2012aj}. The 
other half will then be trapped in the central black hole for $b<b_{crit}$,
or the two individual black holes for $b\gtrsim b_{crit}$ (whose local
Schwarzschild radii would consequently grow by an enormous amount).

A fascinating conjectured aspect of the ultra-relativistic scattering problem is
that it actually does not matter what the source
of the kinetic energy is, be it black holes, or some compact distribution
of matter, such as a neutron star, or even a fundamental particle. 
It is this conjecture behind the arguments that the Large Hadron Collider(LHC)~\cite{Dimopoulos:2001hw,Giddings:2001bu},
or cosmic ray collisions with the Earth's atmosphere~\cite{Feng:2001ib},
could produce black holes in certain extra dimension scenarios
which give a much lower Planck luminosity than our (then erroneous)
4-dimensional analysis predicts. To date, numerical evidence in favor of this
``matter does not matter'' conjecture has only been obtained for a few select
matter models in the head-on collision limit~\cite{Choptuik:2009ww,East:2012mb,Pretorius:2018lfb}.

\subsubsection{The Ringdown}
Due to the uniqueness, or ``no hair'' theorems
of general relativity~\cite{Israel:1967wq,Carter71,Hawking:1971vc,Robinson:1975bv}, the two parameter (mass and angular
momentum) Kerr family of metrics are the only vacuum,
stationary, asymptotically flat black hole solutions without any exterior
(naked) singularities allowed by general relativity in four spacetime dimensions. 
Taken by itself, this would suggest that either black holes are
sets of measure zero and not relevant
to realistic gravitational collapse (the Kerr solutions being axisymmetric and stationary), 
or Kerr black holes are in a sense dynamical attractors
where once a non-symmetric, dynamical horizon forms, evolution causes
the exterior spacetime to ``loose its hair'' and settle down to a Kerr solution.
The latter is a special case of Penrose's {\it final state conjecture}~\cite{PenroseFSC}:
the generic endstate of evolution governed by general relativity, beginning with naked-singularity free 
vacuum initial data on a Cauchy slice of an asymptotically flat spacetime, is
a set black holes flying apart, the local geometry of each approaching that of
a given member of the Kerr family, together with gravitational waves streaming
outward to null infinity. Indeed, this is what seems to generically
happen in gravitational collapses studies and merger simulations to date. In particular
for both quasi-circular inspirals and ultra-relativistic scattering
with $b<b_{crit}$, once a single, common horizon forms the space time
rapidly settles down to a Kerr black hole.
This is accompanied by the emission
of gravitational waves, whose characteristics are largely determined by the quasi-normal
mode oscillation spectrum of the remnant black hole. In analogy with a bell emitting
decaying sound waves after it is hit, this is called the {\it ringdown} of the black hole.
The least damped mode of a Kerr black hole is the $\ell=m=2$ spherical
harmonic mode. The damping rate decreases with the spin of the black hole,
approaching zero for the maximally spinning (extremal) black holes
allowed in general relativity. However, the spins of remnants produced
in comparable mass mergers, as observed by LIGO/Virgo, are sufficiently
far from extremal that their ringdown phases are very short, damping
exponentially with a characteristic e-fold time on order-of-magnitude
the light-crossing time $R_s/c$ of the remnant.

\section{The Gravitational Wave Observational Landscape}\label{sec_obs}

In this section we outline what the current and planned near
future observational campaigns to witness the Universe 
in gravitational waves are. Gravitational wave ``observatories'' 
fall into two categories : those that people have built
specifically for this purpose, and those that the Universe
has fortuitously provided us. The former include 
earlier resonant bar detectors pioneered by Joseph Weber, 
the LIGO/Virgo and Kagra ground based detectors, and various
planned future ground and spacebased detectors. The latter
include a network of millisecond pulsars in our galaxy,
and the cosmic microwave background (CMB).
We will not cover the history of any of these endeavors, instead we
will comment on properties/challenges common to any of them that can be appreciated
with knowledge of the properties of gravitational waves outlined in the
previous section.

Given that general relativity is a theory about the geometric nature
of space and time, and that gravitational waves are propagating
distortions in the geometry, it should not be surprising that essentially
all gravitational wave detectors are composed of elements that are
sensitive to changing distances or times. Moreover, the most
sensitive measurements are those adapted to 
the plus and cross polarized transverse disturbances allowed in general relativity.
This informs the ``L'' shape of the current ground based detectors, that measure
relative changes in distances along the two arms of the detector
through laser interferometry. Pulsar timing relies on the 
remarkably stable rotational periods of certain pulsars, where
models can be built to predict the arrival time of radio pulses
from them to within tens of nanoseconds over a year of observation.
Long wavelength gravitational waves between the earth and the pulsar
will change the arrival times, and the most subtle signals
can be extracted from correlations between changes in arrival
times between pairs of pulsars. Regarding the CMB, this is an image
of the ``surface of last scattering'', where photons were last able
to Thompson-scatter off free electrons (afterward the temperature
of the Universe dropped below a threshold allowing the electrons to
recombined with free protons to form neutral hydrogen). The photons
can pick up a net polarization after Thompson scattering if the
background radiation field is anisotropic. The ability to 
use polarization measurements of the CMB to detect gravitational
waves present then is due to the fact that of the known sources
of anisotropy in the early Universe, only gravitational waves
are able to produce anisotropy that creates a so-called
``B-mode'' polarization pattern over the CMB (as opposed to an ``E-mode'' 
pattern, that both matter anisotropies and gravitational waves can create).

Most sources produce gravitational waves at some characteristic
length or frequency scale. Gravitational wave detectors tend to 
be most sensitive to a frequency/length associated with some scale of the
detector. Therefore, since the different detectors operate
at very different scales, 
they are sensitive to a correspondingly broad spectrum of potential
sources. The ground based detectors are km-scale instruments,
and are most sensitive to physical processes associated
with km-scale sources : stellar mass black holes, neutron stars, and the inner
core of a star undergoing a supernova explosion. 
The space-based LISA instrument is planned to 
be a triangular configuration of satellites with 2.5 million km length arms; 
this is the scale of the smaller of the so-called supermassive black holes
thought to exist in the centers of most galaxies, as well as the
orbits of many close binaries containing white dwarfs, neutron stars and black holes.
Pulsar timing is most sensitive to gravitational waves with periods
close to the years to decades long observation time of the pulsars. This
translates to physical scales on the order of a few light years, and
one of the most promising sources on this scale is an effective
stochastic background from the population of supermassive black hole
binaries in their last stages of inspiral. Gravitational waves
from the early Universe would likely leave a most pronounced
effect on the CMB on scales of order the Hubble radius at the surface
of last scattering, which is roughly $1/1000$ that of the
present day Hubble radius $R_{H_0}\sim 10^{26}$m.

A common problem for all detectors is how weak the gravitational
waves are expected to be when they reach the detectors. This
is true even for the strongest known source---a binary black hole merger---when factoring 
in how far away the event is expected to occur from
Earth. For stellar mass black hole binaries, the observed
merger rate is $\sim 10$ per cubic gigaparsec (Gpc) per year~\cite{LIGOScientific:2021psn}.
In fact, the first event ever detected, GW150914, is still
one of the closest black hole mergers seen to date, at an estimated distance
of $0.4$Gpc $\sim 10^{25}$m. Since gravitational waves decay like 1/distance from
the source, what were metric perturbations of magnitude $h\sim 1/10$ on
the $\sim10^{5}$m scale of GW150914's last orbit, caused
a metric perturbation $h\sim10^{-21}$ as it passed Earth, resulting
in a maximum change in distance along LIGO's $4$km long arms
of $\sim 10^{-17}$m, or about $1/100^{th}$ the diameter of a 
proton!\footnote{Though the 1/distance decay seems like a curse,
and it is for being able to detect rare events
like black hole mergers relatively frequently on a human
timescale, once the tremendous experimental effort needed to cross that threshold
has been met, the 1/distance decay also means it does not take that much more effort to open
up a significantly larger volume of spacetime to observation.
For example, the next (third) generation of ground based detectors
are planned to be about 10 times more sensitive than Advanced
LIGO's design sensitivity. Being able to see 10 times further is enough that 
GW150914-like black hole mergers could be seen throughout the
visible Universe!}
It is not surprising then that one of the most significant
challenges facing all detectors is a thorough understanding and mitigation
of sources of noise that could otherwise swamp, or masquerade as gravitational
waves. This is one of the primary reasons why LIGO 
consists of {\it two} detectors with nearly the same orientation relative
to the sky, but separated by a few thousand kms : a true
gravitational wave must produce signals with similar characteristics
in both detectors, separated in time by at most the few ms of light travel
time between them; conversely, the probability that noise could mimic such
a correlated signal is much less than noise being able to mimic a gravitational
wave in a single detector alone.

A second issue with most gravitational wave detectors is how to interpret
an observed signal once it is confirmed to be of likely astrophysical
origin. Except for the CMB,
the difficulty here is that the signal is a one dimensional time series,
and so these detectors are more akin to seismometers than telescopes
(with the CMB a two dimensional polarization map over the sky can be obtained).
Without theoretical {\it templates} of waveforms from expected sources
to compare against, there is very little other than broad temporal/spectral
characteristics that could be inferred from a novel, or unmodeled source. 
Thus a crucial part of the gravitational wave astronomy endeavor is
to have banks of template waveforms from expected sources. For compact
object mergers, the issue of source interpretation is also
closely tied to detection : current instruments are still not sensitive
enough for the vast majority of mergers to be clearly evident
above the detector noise, and matched filtering is essential
to extract such weak signals from the noise.\footnote{
Matched filtering refers to convolving the detector
signal with a template waveform. If a nearly periodic signal 
with many cycles is present, such as the inspiral phase of a merger,
and an accurate template is phase aligned with
the signal, then with time the convolution will increase the signal
to noise ratio, as the signal will add coherently while typical
noise will not.} This is why solving the two-body problem in general
relativity became such a focused effort within the theoretical
general relativity community beginning in the early 1990's.
Due to the complexities of the Einstein field equations
no analytical solution seems possible, and currently a full solution
(for a given set of orbital parameters) needs to be computed numerically,
which introduces some numerical truncation error.  Moreover,
since numerical solutions are currently too computationally expensive to use to produce
template banks that densely sample parameter space, template banks of practical
use are constructed using various approximation methods; these include
the effective one body (EOB) approach~\cite{Buonanno:1998gg}, modern versions of which
use select numerical results to calibrate the stitching together
of perturbative post-Newtonian inspiral calculations with linear quasi-normal mode ringdown
calculations, and reduced basis models 
constructed from a set of numerical waveforms~\cite{Field:2011mf} (see ~\cite{Barack:2018yly}
for a review of these and other approaches).
In the future
as more sensitive detectors come online, templates will not be needed
as much for detection, though will still be crucial for source
identification and parameter estimation, which
would be hampered if systematic modeling
errors are present in the template libraries.
Thus, even though the first numerical
solution to a general relativistic two body problem describing inspiral, merger and ringdown
was obtained almost two decades
ago~\cite{Pretorius:2005gq}, it is still an active area of research to calculate
ever more accurate binary merger waveforms.

\section{Survey of What Has Been Observed to Date}\label{sec_todate}
In this section we give an overview of the three most important (in our opinion) 
scientific advances to date coming from gravitational wave observation of the 
Universe : testing dynamical strong field gravity, multi-messenger
observation of neutron star mergers, and obtaining the first
glimpses of the demographics of black holes in our Universe. Amongst
the observatories mentioned in the previous section, only LIGO/Virgo
have made actual detections, and we will only comment on these\footnote{
Of course, that is not to say that the absence of a signal does not provide useful information;
e.g. the negative results from the CMB and pulsar timing place constraints
on the magnitude of stochastic backgrounds, and the absence of long lived periodic signals
in LIGO/Virgo data from known pulsars place limits on the size
of quadrupolar deformations (``mountains'') of those neutron stars.}.

{\bf First quantifiable evidence for the existence of black holes as
governed by the theory of general relativity.} Though the evidence
for the existence of black holes has steadily grown since
the first candidates where identified beginning in the 
1960's---the first stellar mass black hole candidate Cygnus X-1, 
the suggested connection between quasars and supermassive black holes, 
our own Milky Way supermassive black hole Sagittarius A$^\ast$---before 
GW150914 the evidence was all circumstantial. In other words,
the only scientifically sound statement one could have made
is that the Universe definitely harbours a few ultracompact objects
and has some unusual sources of electromagnetic emission, and none
of these observations can readily be explained using conventional
physics if Kerr black holes are not involved. 

The gravitational
wave data from black hole mergers is fundamentally different
in this regard, as it is coming from the strong field dynamics
of spacetime itself, and there is already enough signal in some
of the loudest events, such as GW150914, that quantifiable
self-consistency tests can be performed. Most notable
in this regard is the consistency between the inspiral
and ringdown portions of the waveforms. From the inspiral signal
alone an estimate of the progenitor black holes in the
binary can be made, and from this, together with predicted
dynamics of the merger using numerical solutions of the field
equations, the mass and spin of the remnant can be computed. 
From the observed decay and frequency of the ringdown signal alone,
and using black hole perturbation theory calculations,
the mass and spin of the remnant black hole can also be determined.
These two independent measures of properties of the final black hole
must agree if the signal comes from two Kerr black holes colliding 
and forming a remnant Kerr black hole, as described by
general relativity. So far all the LIGO/Virgo data is consistent in this
regard~\cite{LIGOScientific:2016lio,LIGOScientific:2020tif,LIGOScientific:2021sio}, 
albeit the error bars are quite large, as the signal
to noise ratio (SNR) of current events are still quite small
for making precise tests of this kind. As an illustrative
example to put this data and its veracity in context compared to that 
obtained using the Event Horizon Telescope images of
M87, or the Nobel prize winning data of stellar
orbits around Sagittarius A$^\ast$ used to measure its
gravitational mass and confirm its ultracompact nature : we
still cannot rule out that M87 or Sagittarius A$^\ast$ are
ultracompact boson stars\footnote{Boson stars are hypothetical 
star-like objects formed from exotic (i.e., not part of the
standard model of particle physics) self-interacting bosonic matter, in contrast
to neutron stars which are largely composed of fermionic
matter. A boson star's gravitational dynamics is still governed
by general relativity, so it is not an ``alternative'' to a
black hole, but could be a novel class of compact object.};
neither can we exclude
that the {\it progenitors} in GW150914 were ultracompact boson stars.
However for the latter, if they were boson stars, the ringdown 
part of the signal shows they promptly collapsed 
and {\em formed} a Kerr black hole, with mass and angular momentum
consistent with that of the binary just prior to merger.
In other words, even in this hypothetical scenario GW150194 still
gives evidence for the existence of Kerr black holes---exotic 
compact objects more ``bizarre'' than boson stars would need to be invoked
to avoid that conclusion~\cite{Yunes:2016jcc}.

Because of the uniqueness properties of black holes in general relativity,
and if general relativity does accurately describe strong field gravity
on astrophysical scales, then unfortunately we cannot learn
anything more about the {\it physics} of black holes from more
precise merger observations (the {\it astrophysics} of black holes is
a different issued, discussed below). I.e., all black holes
in the Universe are then Kerr black holes to within environmental
perturbations, and perhaps future ultra-precise measurements
of mergers could show imprints of a circumbinary environment,
but there are no novel classes, shapes, or topologies of black holes
to discover. Then, the utility of black hole merger observations for fundamental
physics is essentially entirely to provide detailed tests of
non-linear general relativity as outlined in the previous paragraph.
Of course, as the scientific method requires such tests for the health
of its theories this is a useful endeavor, and we do not need
a motivation other than that. However there is at least
one observationally driven motivation for why one might be
skeptical about the precise nature of
strong field gravity as described by general relativity : dark energy.

On large scales the Universe is observed to be in an epoch of
accelerated expansion; {\it interpreting} this as being due to dark energy
comes from assuming that Einstein gravity accurately describes the geometry of the
Universe on such scales. Specifically, on large scales it is assumed that
with an appropriate time slicing, the spatial metric of the Universe
is nearly homogeneous and isotropic, and its
time evolution is driven by a stress energy tensor characterizing the average energy densities
and pressures of all the matter/energy in the Universe. It is sometimes
stated that today (i.e. away from any ``big bang'' singularities) gravity
on average in the Universe is weak, and certainly on small scales
like our solar system, galaxy, or even that of galaxy clusters
it is weak (except near the rare black hole or neutron star). However, as described
in section \ref{sec_SF}, the strong field regime of general relativity
is not associated with any physical length scale per se, but rather manifests
when some physical scale in the problem becomes commensurate with
the radius of curvature of spacetime. And by that measure, our Universe
is {\it always} in the strong field regime on scales of the Hubble radius $R_H$;
i.e. $H$ is of the same order of magnitude as the Schwarzschild
radius  $R_s\sim\sqrt{3 c^2/8\pi G\rho}$ of a spherical distribution of matter with the same average 
energy density $\rho$ as the matter in the Universe. For example, today
$\rho_0$ is roughly that of six hydrogen atoms per cubic meter, giving $R_{s0}\sim10^{26}$m$\sim R_{H_0}$.
One might complain that the Schwarzschild radius argument does not apply to our
Universe because the latter is not asymptotically flat. Perhaps,
though the point here is not to argue whether or not we are inside a Schwarzschild
black hole, but instead that on scales of the Hubble radius the Universe
must be in the non-linear regime of general relativity
for an entirely different class of solution (the FRLW metrics) to be possible.
Bringing the discussion
back to testing gravity on stellar mass black holes scales, 
if dark energy is telling us general relativity gets things wrong
on the scale of the Hubble radius, we should be cautious
about immediately accepting its predictions for black holes, as
the scale free nature of general relativity implies cosmological
horizons and event horizons reside in a related regime of the 
theory.\footnote{The majority of proposals to explain dark energy
using modified gravity specifically introduce a new physical
length scale into the problem, and if that scale is tuned
to the Hubble radius it would avoid the conclusion
that altering gravity on present day cosmological horizon scales could
have consequences for stellar mass or supermassive black holes.}
The current LIGO/Virgo observations are therefore an important
step toward quantitative verification of the physics of horizons.

{\bf The wealth of knowledge gained from GW170817, the first
binary neutron star merger detected}~\cite{LIGOScientific:2017vwq}.
That so much information was garnered from this event is because
a host of electromagnetic counterpart emission was also seen---the first, and to date 
only gravitational wave-electromagnetic ``multi-messenger'' event~\cite{LIGOScientific:2017ync}. 
Here we briefly comment on the highlights.
The first is that a short Gamma Ray Burst (sGRB) was detected $\sim 1.7$s after
the observed gravitational wave inspiral, the latter
which ended a few ms before the presumed collision of the two
neutron stars (this, and any post-collision gravitational waves
where not seen by LIGO/Virgo, which is as expected as they occur
at frequencies several times higher than what LIGO/Virgo is sensitive to).
The origin of sGRBs has long been a mystery, though one of the leading
hypothesis for their formation is they are produced in polar jets powered by
accretion onto the remnant of a binary neutron star merger (whether
it be a hypermassive neutron star or a black hole that formed,
though the latter seems to be a more favorable environment for jet formation).
The coincidence of the gravitational wave emission and sGRB, both
in terms of time and region of the sky where both fluxes
appeared to come from, gives the first solid evidence that at least a class
of sGRBs are produced following a binary neutron star merger.
Assuming this connection, together with the estimated distance to the event of $40$Mpc,
then also gives a direct measurement of the speed of gravitational
waves relative to the speed of light, and a remarkably tight constraint  
for a first measurement : the two speeds are the same to within approximately
1 part in $10^{15}$~\cite{LIGOScientific:2017zic}.  

Almost immediately after GW170817 was detected, a world-wide effort
was undertaken by astronomers to search for other electromagnetic counterparts,
and within 11 hours a bright but fading optical transient was identified in the galaxy NGC 4993.
Follow up observation over the subsequent weeks saw the event in radio, X-ray, infrared
and the ultraviolet. The observed properties of the emission are consistent with the 
neutron star merger having produce a so-called kilonova (or macronova)~\cite{Kasen:2017sxr}.
During merger, a small fraction ($\sim0.1-1\%$) of the neutron star's material is tidally 
ejected from the system at mildly relativistic speeds ($\sim c/3$), and over
the subsequent few seconds following merger a similar amount of material
can be blown away from a hot accretion disk formed around the remnant, at similar
but slightly lower velocities. This initially high density material is very
neutron rich, and as it expands heavy elements (with atomic number in the
range $Z\in28..90$) are formed through the r-process. Many of these elements 
are radioactive with relatively short lifetimes, and it is their decay 
over the subsequent days that produces the light of the kilonova. This also
confirms that neutron star mergers are one of the sites where a significant
fraction of the Universe's heavy elements are produced---it is quite
likely that the gold and platinum we humans so love to adorn ourselves with
are the ashes of ancient galactic neutron star mergers.

The late stages of the gravitational wave emission in GW170817 also
showed mild deviation from the predictions of a black hole inspiral,
indicative of tidal deformations occurring in both neutron
stars. The strength of the tidal deformation is governed by the equation
of state of matter at nuclear density, which is not theoretically
well understood today, nor accessible to experiments on Earth to investigate.
Thus neutron star mergers offer an avenue to explore this extreme state
of matter, and though this first event did not provide strong constraints
on competing models, this is one of the subjects future observations
are expected to bring increasing clarity to.

Another subject that GW170817 allowed gravitational wave astronomy to
take a first step in, but will also require more future observations
to make a useful contribution toward, is measuring the local expansion
rate of the Universe. This is typically done by measuring both the distance $d$
and redshift $z$ to a set of sources in galaxies, and the expansion history
can be inferred from the relationship $z(d)$ (for small redshifts,
so nearby galaxies, $z \approx H_0 d/c$, where $H_0$ is the Hubble constant). Measuring the distance
to a source is quite challenging. One method relies on a so-called standard
candle, where the intrinsic luminosity $L$ of a source is assumed known,
and hence the observed flux is simply $L/4\pi d^2$. Type Ia supernovae
are the most well known standard candles, though inferring their
intrinsic luminosity relies on several calibration steps, including
the cosmic distance ladder. With a binary neutron star merger where
a counterpart is seen (and hence the host galaxy identified for a redshift measurement), a
luminosity distance-redshift measurement can be obtained that bypasses
all of these calibration steps, since the intrinsic luminosity of the
merger is known from the general relativity waveform calculation.
This makes a binary neutron star merger a standard ``siren'' (siren is used here
instead of candle as the last stages of inspiral emit waves in the audio
frequency range).\footnote{Binary black holes are also standard sirens, and better
ones in fact, as some uncertainty will be present in the neutron
star measurements until the nuclear equation of state is known. However
binary black holes are not typically expected to be in an
environment where a strong electromagnetic counterpart will be produced,
and none have been observed to date.} GW170817 has already by itself
allowed a measurement of $H_0$ to within about $10\%$; though this is not
an improvement over other existing measurements, the more multimessenger
binary neutron star events that are observed, the tighter the standard siren based value
will become. Eventually, this might prove to be instrumental to help resolve
the present ``Hubble tension'' : measurements of $H_0$ inferred by the Planck satellite's
observation of the CMB show a small, but statistically significant mismatch with $H_0$ 
obtained using supernovae data (see e.g.~\cite{Freedman:2021ahq}). 

{\bf Tentative hints pointing to an ``unusual'' stellar mass black hole population.}
Of the almost 100 signals LIGO/Virgo have so far detected that are of
likely astrophysical origin, the vast majority are consistent
with binary black hole merger 
templates~\cite{LIGOScientific:2018mvr,LIGOScientific:2021usb,LIGOScientific:2021djp}.\footnote{The 
remainder also match binary black hole templates,
but when one or both companions have masses less than $\sim 2.5 M_\odot$,
the event is classified as a black hole-neutron star or binary neutron star
merger respectively. To be able to distinguish between black holes
and neutron stars from the gravitational waves alone would require
observation of the higher frequency late stages of inspiral/merger,
or a high enough SNR event that
the effect of tidal deformation is already evident in the earlier lower frequency
inspiral that can be observed with present detectors.}
As discussed above, if general relativity is correct, then we know
these are all merging Kerr black holes, forming remnant Kerr black holes. The
utility then in having this large number of events, and anticipating
even more in the years to come, is to learn what the distribution of masses
and spins of this sub-population of black holes in the Universe is as a function
of time (redshift). This will provide information
on the fates of the most massive stars that are expected to form
black holes at the ends of their lives, as well as binary
formation channels. Regarding the latter, the two thought to be predominant
are from stellar binaries where both stars are massive enough
to form black holes, and dynamical assembly in dense cluster
environments (following chance encounters between either two isolated black
holes, a binary containing a black hole and a single black
hole, or a binary-binary interaction where each contains a black hole). 
Though even 100 events is not yet enough to give definitive
answers to some of these population questions, there are already
some interesting trends, and a few outliers that 
are somewhat puzzling or surprising (at least without hindsight
to select amongst the many reasonable arguments present
in the prior body of literature speculating about the unknown).

The first surprise came with GW150914, in that both progenitor
black holes had masses ($\sim$ $29 M_\odot$ and $36M_\odot$)
at least twice that of any known stellar mass black hole candidate
in the Milky Way (see e.g. \cite{Casares:2017jah}). Subsequent
detections showed that GW150914 is not an outlier in this regard,
and most (though not all) LIGO/Virgo black hole progenitors
are more massive than known galactic black holes. This could
partly be a selection effect, as LIGO/Virgo is 
more sensitive to higher mass mergers, and also that
the X-ray binary systems that have been used to identify
galactic black holes might be a distinct population of binaries
from those that lead to black holes that merge within a Hubble time.

A second puzzle is that the vast majority of progenitor
black holes seem to have very low spin (the remnants 
acquire higher spin, around $60-80\%$ that of the maximum
allowed for Kerr black holes). Or to be more technically precise,
given the detector's current sensitivities, with most inspirals
a confident measurement can only be made of the net spin angular momentum aligned
with the orbital angular momentum---for most mergers detected
to date this result is consistent with zero (to within error bars). 
There are three primary configurations that can achieve this : (1) the individual
black holes actually do have close to zero spins, (2) the individual
black holes have roughly equal but opposite spin angular momenta,
one aligned, the other anti-aligned with the orbital angular momentum, (3) the black holes
have arbitrary spins (less than extremal) but the spin vectors
are mostly {\it within} the orbital plane. Both options (2) and (3) are difficult
to explain with a binary formed from a stellar binary, where one
would typically expect the spin vectors to be almost aligned with the orbital
angular momentum vector. Options (2) and (3) are consistent with the occasional
dynamically assembled binary, as there is no preferential orientation
for an essentially random close encounter, but one would not expect
this for the majority of events as currently observed. Thus (1)
seems the most plausible explanation at the moment. Given how challenging
it is to simulate stellar collapse at present, hence have 
robust predictions for what the initial spin distributions of
black holes should be, the observations will serve as useful guide posts
for ongoing theoretical studies of collapse. 

There are more
speculative suggestions for why the progenitors have low spin. One is that
many of these low-spinning black holes are primordial
in nature, meaning the black holes might have formed at a very early
epoch in the Universe (well before structure formation) from rare super-high
density fluctuations in the background radiation field. The concrete
mechanisms people have proposed for this typically produce very
low spin black holes (see e.g.~\cite{Carr:2021bzv} for a review). Another possibility is that there
are as of yet undiscovered ``ultra-light'' particles, with Compton 
wavelengths on the order of the tens of kilometer scale of the Schwarzschild
radii of stellar mass black holes. Such particles can form bound states
around the black holes, and if the black hole is spinning, these bound states can
grow by a so-called superradiant interaction with the surrounding spacetime~\cite{Brito:2015oca}. 
In reaction, the black hole spins down, possibly quite rapidly on astrophysical 
timescales
(much less than the relevant gigayear timescale, which is order of magnitude the maximum time 
between a black hole's birth and when it should suffer a collision
with another to be visible to LIGO/Virgo). Of course, even if such 
particles exists, they might not be the reason for the low spin black
hole population --- that could still just be due to properties of stellar collapse
and black hole formation.

The third surprise relates to several outlier events,
the two most prominent being GW190521 and GW190814,
that seem to have progenitor compact objects in the so-called ``mass gaps''. 
GW190814 is the merger of a $\sim 23\pm 1 M_{\odot}$ (presumed) black hole
with a $\sim 2.6\pm0.1 M_{\odot}$ compact object~\cite{LIGOScientific:2020zkf}. 
GW190521 is the merger of a $\sim 85\pm20 M_{\odot}$ black hole
with a $\sim 66\pm18 M_{\odot}$ black hole~\cite{LIGOScientific:2020iuh}.
Regarding GW190814, arguments from stellar collapse 
studies, as well as a dearth
of candidates from our known galactic compact object population,
suggest objects with masses in the range $\sim 2.5 M_{\odot} - 5 M_{\odot}$
do not typically form in stellar collapse. Moreover, it is currently
unknown if the maximum allowed mass for a neutron star
can reach $2.5 M_{\odot}$; if it turns out to be less
than $2.5 M_{\odot}$, the lower mass companion of 
GW190814 would be challenging to explain (or be an exceedingly
rare object, for example a low mass black hole formed via a prior binary neutron
star merger). Regarding GW190521, 
stellar structure theory suggests stars with cores in the 
mass range $\sim 65 M_{\odot} - 135 M_{\odot}$ are 
subject to the so-called pulsational pair-instability supernova processes,
which blows the cores apart leaving behind no remnant.
However, similar to the issue
of the spin of a black hole at birth, there is a fair amount
of uncertainty to the exact range of this mass gap,
and given the error bars in the mass measurements, 
there is only mild tension between GW190814
and conventional theories.

\section{The Future of Gravitational Wave Astronomy}\label{sec_future}

Einstein's theory of general relativity is over 100 years old, 
and the quest to observe the Universe in gravitational waves is over 
50 years old, beginning with Joseph Weber's pioneering attempts
in the 1960's. Despite these long histories, the field
of gravitational wave astronomy is in its infancy, with the first
detection only 6 years ago. Though many signals observed to date
are solidly above the threshold for confident
assertion that they are gravitational waves coming
from astrophysical sources, they are still not loud enough
for high precision tests of strong field gravity, nor for high accuracy estimation
of all source parameters. Moreover, most of these detections
have relied on theoretical templates of expected sources, which 
improves the effective sensitivity of the detectors. Thus any truly
novel source will likely only
be discovered once the detector sensitivities are well above
the threshold the new source could otherwise have been seen using templates.
The one exception here is a source that emits a short burst well approximated
by a sine-Gaussian, as LIGO/Virgo do employ searches using such templates
(this can be thought of as an ``unmodeled'' search in the sense that
there is no particular astrophysical source from which the template
is derived). 

To realize a future where a detailed picture of the Universe
in gravitational waves is attained will thus require more sensitive detectors
that cover a broader range of frequencies than at present. These are being
planned, and within the next decade or two we can expect an order of magnitude
improvement over essentially the entire slate of observational campaigns. 
LIGO is within a factor of two of the original ``Advanced LIGO'' design sensitivity,
which should be reached during the next observing campaign (beginning
late 2022-early 2023), when
the KAGRA detector in Japan will also join the LIGO/Virgo network~\cite{KAGRA:2013rdx}.
Following that, the plan is for an ``A+'' upgrade that will improve
sensitivity by another factor of two, and LIGO India will join the network
(anticipated to start in 2025). To improve sensitivities significantly
beyond this will require new facilities, and several third generation
designs are being planned for the 2030's, including Cosmic Explorer and the Einstein
Telescope~\cite{Kalogera:2021bya}. These could further increase sensitivity by a factor
of 10, as well as offer improved frequency bandwidth over both lower (earlier in the inspiral for
binary compact objects) and higher frequencies (merger regime for binary neutron stars).
New technologies are also being considered, most promising among these are
atom interferometers~\cite{Geiger:2016cme}, though it is less clear what the timeline
for their deployment is. The spacebased LISA mission is expected
to launch in the late 2030's. Both LISA and third generation ground based
detectors could see black hole mergers with SNR
close to a thousand (the current SNR record holder is GW170817, at $\sim 32$).
CMB measurements of B-mode polarization over the next decade (e.g. with the
Simons Observatory~\cite{SimonsObservatory:2018koc} currently under construction, and the LiteBIRD
satellite planned to be launched by the end of the decade~\cite{LiteBIRD:2020tzb}),
should lower the threshold above which cosmic gravitational waves would be
observed by about an order of magnitude.
The sensitivity of the pulsar timing network 
increases roughly with the square-root of the observation time,
and could be accelerated with the discovery of more  
highly stable pulsars clocks to add to the network (see e.g.~\cite{Hobbs:2017oam}).

We conclude with a brief discussion of what we can hope/expect to learn from these observatories
if everything goes according to plan. At the very least we can expect
an ever clearer picture of the demographics of compact objects
in our Universe unfolding, improved tests of the dynamical
strong field regime of general relativity, tighter constraints
on the Hubble constant $H_0$ from gravitational wave standard sirens,
first detection of a stochastic
background of gravitational waves from unresolved supermassive black hole binaries, and either
a first measurement of a primordial gravitational wave background
from an inflationary epoch in the early Universe, or a bound
on the latter that would severely challenge the inflationary paradigm.
If we are fortunate, a binary neutron star merger as close or closer than
GW170817 will occur during the era of the third generation of ground based
detectors, which would provide unprecedented insight into the nature
of matter at the extreme nuclear densities present in the interior
of neutron stars. If we are very fortunate, a star will go supernova (while the detectors are on!)
in our neighborhood of the Milky Way, 
which should be close enough for us
to be able to hear it in gravitational waves.

A wish opening up our view of the Universe to the medium of gravitational
waves has always been that new, unexpected and surprising sources
will be discovered. Though of course we cannot make a list of the 
truly unexpected, there are sources that people have speculated
about that would be surprising, and some quite revolutionary, if discovered.
These include cosmic strings, ultralight particles driving black hole superradiance,
new kinds of compact objects such as boson stars, and various ``exotic'' horizonless compact
object alternatives to black holes. The latter include fuzzballs, gravastars
and AdS (Anti de Sitter) black bubbles, all inspired by ideas on how ``quantum gravity''
could resolve the singularities of general relativity and apparent information loss paradox
associated with black holes that evaporate via the Hawking process.
But perhaps the biggest surprise of all would be if, once all is said and done, there
are no surprises beyond a few black holes having been born with their two strands of Kerr hair
standing mildly out of place.


\vspace{0.1in}
{\bf Funding}

The author acknowledges support from NSF Grant No.
PHY-1912171, the Simons Foundation, and the Canadian Institute For Advanced
Research (CIFAR).


\bibliographystyle{hunsrt}
\bibliography{references}

\end{document}